\documentclass[11pt,twoside]{article}
\usepackage{asp2010}

\resetcounters

\begin{document}

\title{Episodic accretion, radiative feedback, and their role in low-mass star formation}
\author{Dimitris Stamatellos$^1$, David Hubber $^{2,3}$, and Anthony Whitworth$^1$
\affil{$^1$School of Physics \& Astronomy, Cardiff University, 5 The Parade, Cardiff CF24 3AA, UK}
\affil{$^2$Department of Physics \& Astronomy, University of Sheffield, Hounsfield Road, Sheffield S3 7RH, UK}}
\affil{$^3$School of Physics and Astronomy, University of Leeds, Leeds LS2 9JT, UK}

\begin{abstract}
It is speculated that the accretion of material onto young protostars is episodic. We present a computational method to include the effects of episodic accretion in radiation hydrodynamic simulations of star formation. We find that during accretion events protostars are ``switched on", heating and stabilising the discs around them. However, these events typically last only a few hundred years, whereas the intervals in between them may last for a few thousand years. During these intervals the protostars are effectively ``switched off", allowing gravitational instabilities to develop in their discs and induce fragmentation. Thus, episodic accretion promotes disc fragmentation, enabling the formation of low-mass stars, brown dwarfs and planetary-mass objects. The frequency and the duration of episodic accretion events may be responsible for the low-mass end of the IMF, i.e. for more than 60\% of all stars.
\end{abstract}

\section{Introduction}

Young protostars grow in mass mainly by accreting material from their surrounding discs. The accretion is speculated to be episodic \citep{Herbig77,Dopita78, Reipurth89,Kenyon90,Evans09,Dunham10}, as e.g. in  FU Ori-type outbursts in which a large fraction of the star's mass is delivered within a few hundred years \citep{Hartmann96,Zhu10}.  The estimated accretion rates during these bursts are up to  $\sim 5\times 10^{-4}\,{\rm M}_{\sun}\,{\rm yr^{-1}}$, so each burst can deliver a few $10^{-2}\,{\rm M}_{\sun} $ onto the protostar.  Further evidence for episodic accretion comes from the periodically spaced knots seen in bipolar jets \citep[e.g.][]{Reipurth89}. Bipolar jets are driven off along the protostellar rotation axis by the energy released as material spirals in and accretes onto the protostar; periodically spaced knots therefore imply episodic accretion. Finally, the luminosity problem provides indirect observational support for episodic accretion. By the end of the Class 0 phase ($\sim10^5$ yr), half the protostar's final mass has been accumulated. Thus for a final mass of $1\,{\rm M}_\odot$, the mean accretion rate onto the protostar must be $\sim5\times 10^{-6}\,{\rm M}_{\sun}\,{\rm yr}^{-1}$, and the mean accretion luminosity must be $\sim 25\,{\rm L}_{\sun}$. This is much larger than the observed bolometric luminosities of typical solar-type protostars \citep{Kenyon90,Evans09}. Episodic accretion mitigates this problem, as the luminosity is only large intermittently, and most protostars are observed between bursts \citep{Dunham10,Offner11}.

Accretion of material onto a young protostar results in large amounts of accretion luminosity which may effect the evolution of the protostar's surrounding disc \citep{Krumholz06,Bate09,Offner09,Urban10,Krumholz10, Offner10}. This is particularly important for the low-mass end of the IMF (e.g. for low-mass stars and brown dwarfs) as gravitational fragmentation of massive extended discs is considered to be one of the main ways of producing these objects \citep[e.g.][]{Whitworth06,Stamatellos09,Stamatellos09b, Boley10, Kratter11}. Thus, in order to determine the full effect of radiative feedback in low-mass star formation, it is important to describe in detail the role of episodic accretion.

\section{An episodic accretion model for hydrodynamic simulations}
Material spirals from the outer disc region towards the young protostar due to the work of the gravitational instabilities in transporting angular momentum outwards in the disc. Once this material reaches within a few AU from the protostar  the GIs becomes ineffective as the disc is hotter there. The material is then deposited on a notional inner accretion disc (IAD), where it piles up until it becomes hot enough that thermal ionisation couples the matter to the magnetic field. At this point the magneto-rotational instability (MRI) is activated, transporting angular momentum outwards, and thereby allowing the matter accumulated in the IAD to spiral inwards and onto the central protostar.

 Simulations of self-gravitating hydrodynamics on the scale of molecular cloud cores (i.e. sizes from $10^4$ to $10^5\,{\rm AU}$ and masses from $0.3$ to $10\,{\rm M}_\odot$) can achieve sufficient resolution to capture the formation of discs around young protostars, and the effects of GIs in the outer regions of such discs. However, such simulations cannot capture what happens in the inner disc region, where sinks are invoked. Contrary to previous studies \citep[e.g.][] {Attwood09}, we assume that the matter which enters a sink is not instantly accreted onto to protostar but it is accumulated on the IAD, within the sink radius.  We adapt the model of  \cite{Zhu10} to phenomenologically describe the properties of episodic accretion. The details of the model are discussed in \cite{Stamatellos11}.   An episodic accretion event is initiated when 
\begin{equation}
M_{\rm IAD}\geqslant M_{\rm MRI}\ \, 
\end{equation}
which is presumed to correspond to an IAD temperature of $T_M=1400\ {\rm K}$. In the above equation
\begin{equation}
M_{_{\rm MRI}}\simeq 0.13\,{\rm M}_\odot\,\left(\frac{M_\star}{0.2\,{\rm M}_\odot}\right)^{2/3}\,\left(\frac{\dot{M}_{_{\rm IAD}}}{10^{-5}\,{\rm M}_\odot\,{\rm yr}^{-1}}\right)^{1/9}\,,
\end{equation}
is the mass accreted onto the protostar in an outburst event. The duration of the event is set as
\begin{equation}
\Delta t_{_{\rm MRI}}\simeq 250\,{\rm yr}\,\left(\frac{\alpha_{_{\rm MRI}}}{0.1}\right)^{-1}\,\left(\frac{M_{_{\rm MRI}}}{0.13\,{\rm M}_\odot}\right)\,,
\end{equation}
where $\alpha_{\rm MRI}$ is the viscosity parameter provided by the MRI. To match observations of FU Ori stars, we assume an exponential decay of the accretion rate 
\begin{equation}
\dot{M}_{\star{\rm,EA}}=\frac{M_{\rm MRI}} {\Delta t_{\rm MRI}}\,e^{-\frac{(t-t_0)}{\Delta t_{\rm MRI}}}\,,\ t_0<t<t_0+\Delta t_{\rm MRI}\,.
\end{equation}

Once the accretion rate of the protostar has been determined using the above model, its luminosity at any time is given by
\begin{equation}
L=\left( \frac{M}{{\rm M}_{\sun}}\right)^3{\rm L}_{\sun}+\frac{f G M \dot{M}}{R}\,,
\end{equation}
where $M$ is the mass of the protostar, $R$ its radius, and $\dot{M}$ is the accretion rate onto it. The first term on the righthand side is the intrinsic luminosity of the protostar (due to contraction and nuclear reactions in its interior). The second term is the accretion luminosity. $f=0.75$ is the fraction of the accretion energy that is radiated away at the photosphere of the protostar, rather than being expended driving jets and/or winds \citep{Offner09}. We assume $R=3{\rm R}_{\sun}$ is the typical radius of a young protostar \citep{Palla93}. At the initial stages of star formation the accretion luminosity dominates over the intrinsic  luminosity of the protostar.

The above model describes macroscopically the effects of episodic accretion, which  in turn regulates the radiative feedback from the protostar.

\section{The role of episodic accretion in low-mass star formation}

\begin{figure}
\centerline{
\includegraphics[height=11.5cm,angle=-90]{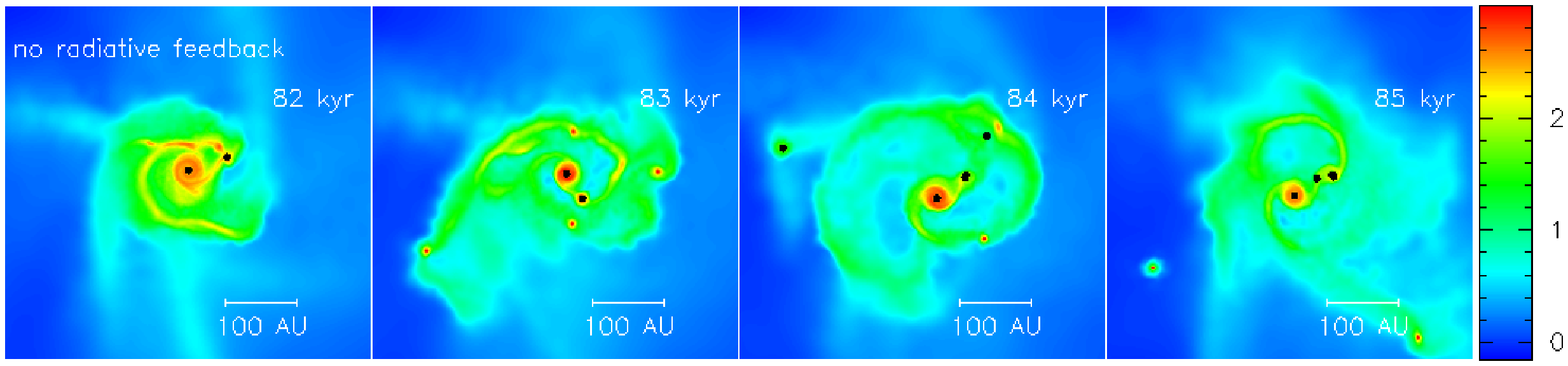}}
\caption{Evolution of the accretion disc around the primary protostar forming in a collapsing turbulent molecular cloud core, without radiative feedback from the protostar. The disc around the primary protostar increases in mass, becomes gravitationally unstable, and fragments to form 3 low-mass stars, 2 brown dwarfs and 2 planetary-mass objects.  The colour encodes the logarithm of column density, in ${\rm g\ cm}^{-2}$. }
\label{fig1}
\centerline{
\includegraphics[height=11.5cm,angle=-90]{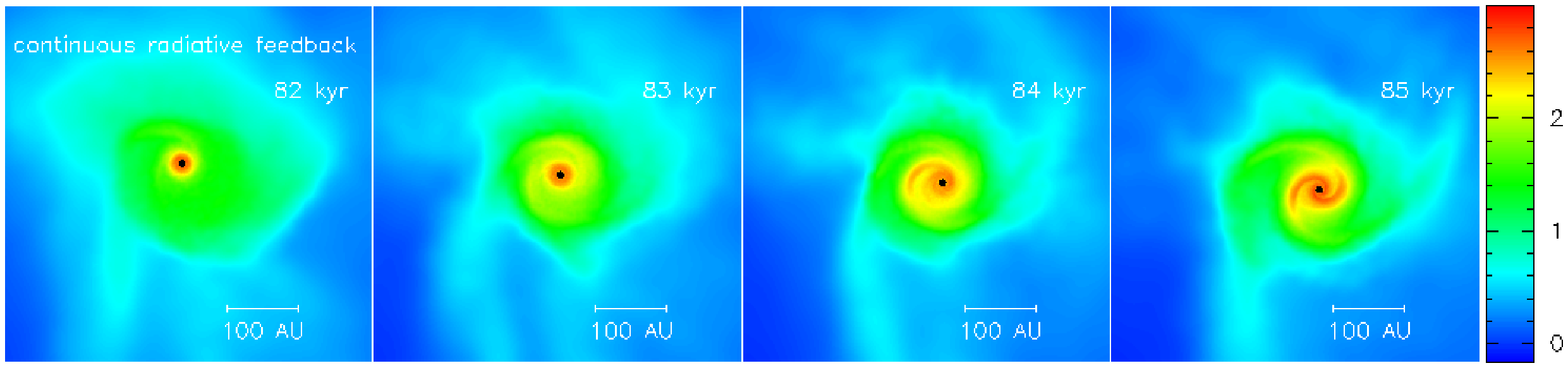}}
\caption{Evolution of the accretion disc around the primary protostar with  continuous accretion, continuous radiative feedback from the protostar. The disc grows in mass, but radiative feedback makes it so hot that it does not fragment.}
\label{fig2}
\centerline{
\includegraphics[height=11.5cm,angle=-90]{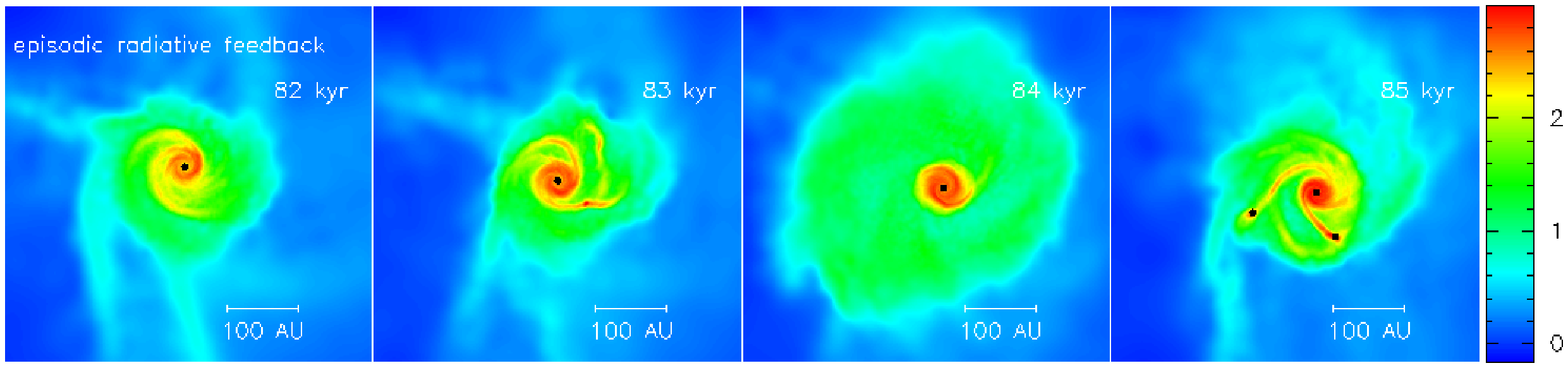}}
\caption{Evolution of the accretion disc around the primary protostar with episodic accretion and episodic radiative feedback from the protostar. The disc becomes gravitationally unstable (first and second column) but fragmentation is damped by heating due to an accretion burst (third column); however, after this burst the disc cools sufficiently to undergo gravitational fragmentation; 2 low-mass stars form in the disc.}
\label{fig3}
\end{figure}

To evaluate the consequences of episodic accretion for disc fragmentation and low-mass star formation, we perform radiation hydrodynamic simulations of a collapsing turbulent molecular core.  We use the SPH code SEREN \citep{Hubber11}, in which the radiative transfer is treated with the diffusion approximation of \cite{Stamatellos07}. The outer envelope of the core extends to $5\times10^4$~AU,  and the total core mass is 5.4~${\rm M}_{\odot}$. We perform three simulations, all with the same initial conditions, and differing only in their treatments of the luminosities of protostars \citep[see][for details]{Stamatellos11}.  In all three simulations the initial collapse leads to the formation of a primary protostar (i.e. a first sink) at $t\sim77$ kyr, and this quickly acquires an extended accretion disc (Figs.~\ref{fig1}-\ref{fig3}). The simulations only diverge after this juncture.

%%%%%%%%%%%%%%%%%%%%%%%%%%%%%%%%%%%%%%%%%%
 %%%%%%%%%%%%%%%%%%%%%%%%%%%%%%%%%%%%%%%%%%

\begin{figure}
\centerline{
\includegraphics[height=11cm,angle=-90]{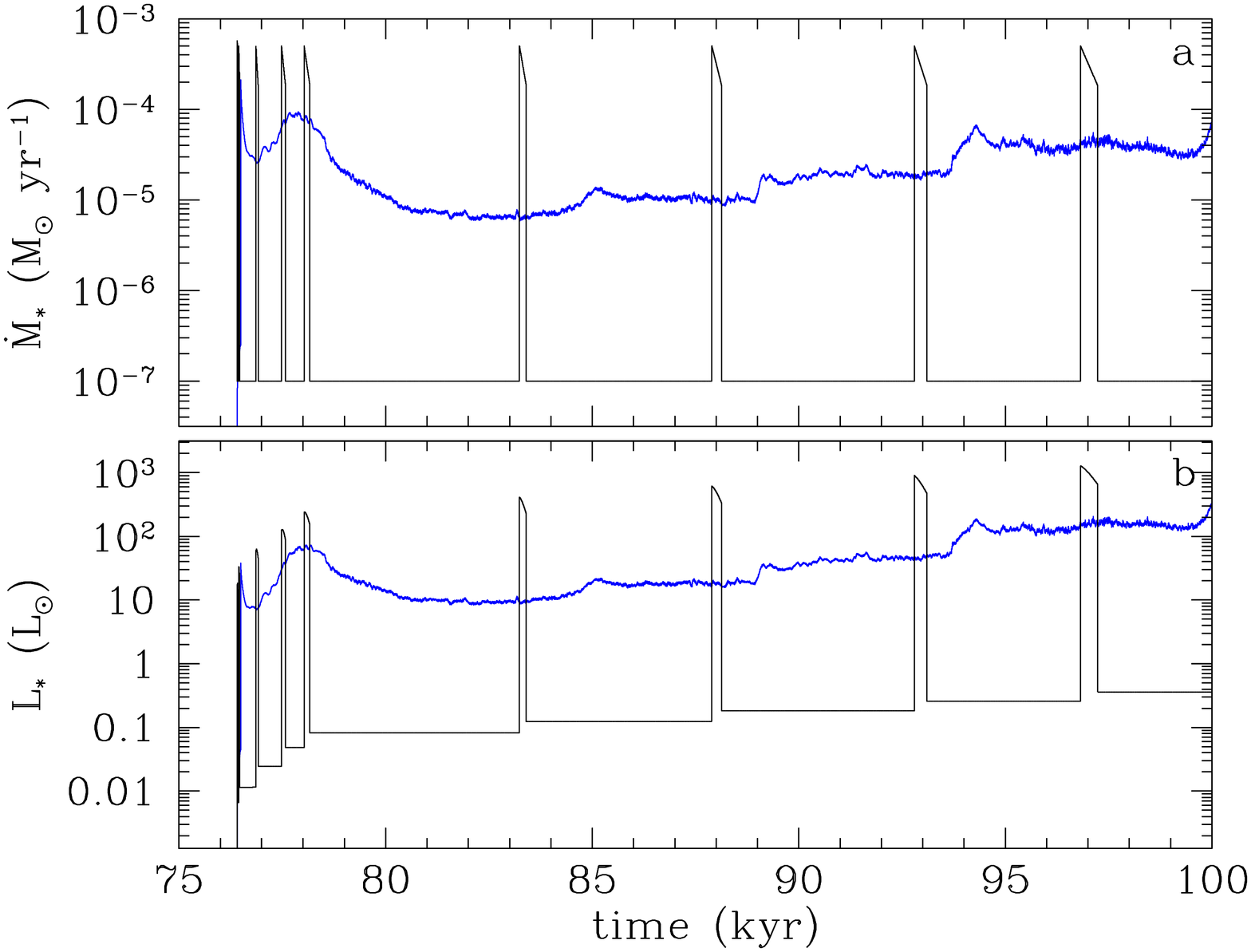}}
\caption{The evolution of the accretion rate and the luminosity of primary protostar. Blue lines correspond to the  the case with continuous accretion and feedback, and black lines for the case with episodic accretion and feedback. (a) shows the accretion rate onto the primary protostar amd (b) shows the accretion luminosities, for the cases with continuous (blue) and episodic (black) accretion.}
\label{fig4}
\end{figure}

\paragraph{No radiative feedback from protostars}
In the first simulation there is no radiative feedback from the protostar, and the gas in the disc is only heated by compression and by viscous dissipation in shocks. The disc stays cool and fragments resulting in the formation of 7 secondary protostars, with masses from 0.008~${\rm M}_{\odot}$ to 0.24~${\rm M}_{\odot}$. This is illustrated in Fig.~\ref{fig1}.

\paragraph{Continuous accretion/continuous radiative feedback from protostars}
In the second simulation, we assume that the matter entering a sink is immediately accreted onto the protostar at its centre. The accretion of material onto the protostar is continuous (see. Fig.~\ref{fig4}, top, blue line). This results in an accretion luminosity which is typically 10 to 100 L$_{\odot}$ (see Fig.~\ref{fig4}, bottom, blue line). It therefore heats the surrounding disc, entirely suppressing disc fragmentation, at all radii, as it is illustrated in Fig.~\ref{fig2}.

\paragraph{Episodic accretion/episodic radiative feedback from protostars}
In the third simulation, the accretion rate into the protostar during the MRI-driven accretion bursts is very high (see Fig.~\ref{fig4}, top, black line), resulting in high luminosity (see Fig.~\ref{fig4}, bottom, black line). This heats and stabilises the disc against fragmentation. However, as the mass of the protostar builds up, the mass that is required to activate MRI also goes up, and at the same time the accretion rate into the sink goes down, so that the intervals between bursts become longer. During these intervals the luminosity of the protostar is relatively low, hence the disc cools down. Within a few kyr of the formation of the primary protostar, the interval between successive accretion bursts has increased to $\sim 5$~kyr, and this is sufficient time to allow the outer disc at $R\sim50-150$~AU, to undergo gravitational fragmentation.
This is illustrated on Fig.~\ref{fig3}. In the first two frames, at 82 and 83 kyr, the disc is unstable and tries to fragment, but then at 84 kyr it is heated by an accretion outburst and stabilised. Once the outburst is over, the disc cools back down fast, and fragments to produce 2 low-mass hydrogen-burning secondaries. 

\section{Conclusions}
If episodic accretion is a common phenomenon among young protostars as observational and theoretical evidence suggests \citep{Herbig77,Dopita78, Reipurth89, Hartmann96,Greene08,Evans09,Peneva10,Dunham10}, it may limit the effect of the luminosity of a protostar on its environment and create favourable conditions for disc fragmentation to occur. As disc fragmentation is predominantly a mechanism for forming low-mass stars and brown-dwarfs, episodic accretion may then have a significant influence on the lower end of the IMF \citep{Stamatellos07b,Stamatellos09,Stamatellos09b, Stamatellos11b}.

\end{document}